# Framework for 3D TransRectal Ultrasound (TRUS) Image-Based Tracking
## - Example of Use –
## Evaluation of 2D TRUS Prostate Biopsies Mapping

Mozer P [1,2], Baumann M [1,3], Chevreau G [1,2], Daanen V [3],
Moreau-Gaudry A [1], Troccaz J [1].

[1] University J. Fourier, TIMC Laboratory, Grenoble, France; CNRS, UMR 5525
[2] Urology Departement, Pitié-Salpétrière Hospital, Paris, France
[3] Koelis, SAS, La Tronche, France.

## *Introduction and Objective:*

Prostate biopsies are mainly performed under 2D TransRectal UltraSound (TRUS) control by sampling the prostate according to a predefined pattern. In case of first biopsies, this pattern follows a random systematic plan. Sometimes, repeat biopsies can be needed to target regions unsampled by previous biopsies or resample critical regions (for example in case of cancer expectant management or previous prostatic intraepithelial neoplasia findings). From a clinical point of view, it could be useful to control the 3D spatial distribution of theses biopsies inside the prostate.

Modern 3D-TRUS probes allow acquiring high-quality volumes of the prostate in few seconds. We developed a framework to track the prostate in 3D TRUS images. It means that if one acquires a reference volume at the beginning of the session and another during each biopsy, it is possible to determine the relationship between the prostate in the reference and the others volumes by aligning images. We used this tool to evaluate the ability of a single operator (a young urologist assistant professor) to perform a pattern of 12 biopsies under 2D TRUS guidance.

## *Methods:*

After approval by ethical committee, the operator performed 12-core TRUS biopsies on 32 patients according to a classical pattern (hence 12 identical square coronal sectors – Figure 1) using a 2D-3D TRUS probe (RIC5-9 on a Voluson-i, both from General Electric Medical). Prostate volume was on average 45ml (min: 20ml; max: 100ml).

- During the biopsy session:
    - Before the first biopsy, a 3D reference volume is acquired.
    - For biopsy targeting, the probe is switched to 2D mode.
    - After each biopsy gun shot, the needle is left inside the prostate, on average during 5 seconds, and a 3D TRUS volume is acquired. During this acquisition the operator took care to apply a minimal force on the probe to minimize deformation on the prostate.
- After the biopsy session:

The needle is manually selected in each 3D volume and is fused automatically into the reference volume with our rigid image-based registration algorithm [1]. Registration success was determined visually. To validate registration, we segmented clearly visible point-like fiducials (e.g. calcifications) in the volumes. The distances between corresponding fiducials after application of the registration transformation were used as gold standard for accuracy evaluation.

After registration, biopsies can be represented into the reference volume (Figure 2 and Figure 3) allowing an analysis of their spatial distribution. To perform a quantitative analysis, the 3D volume is reformatted in the coronal plane and the preoperative targets are created (Figure 4). For each target we computed the percentage of planned biopsies hitting the target and the average biopsy length inside the target. Nevertheless, as the apex lateral target area was often small on the preoperative planning, we fused it with the apex parasagital target for the analysis.

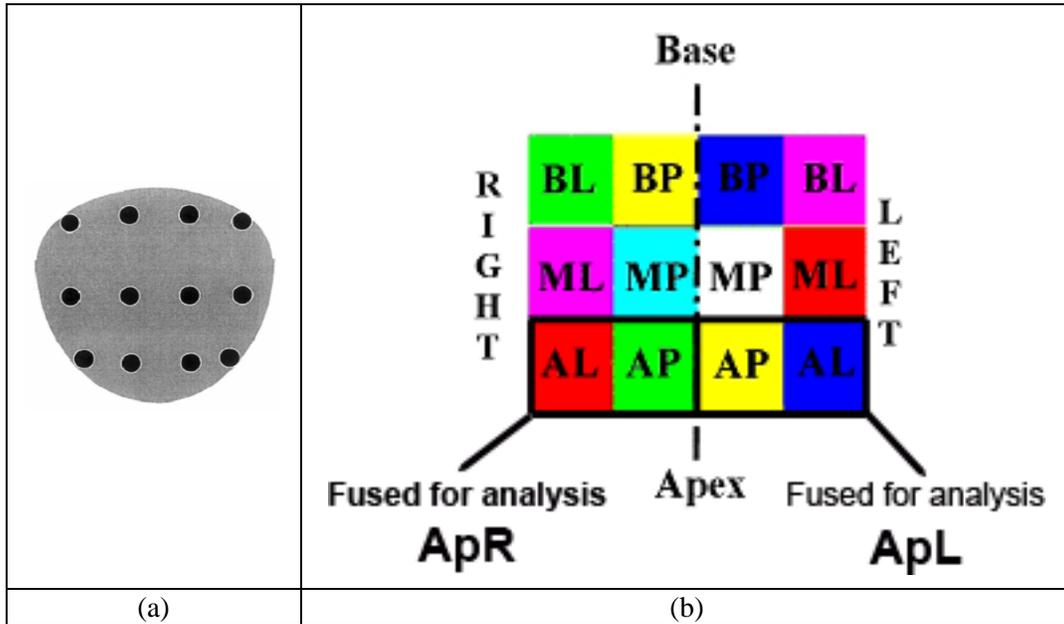

(a) (b)

**Figure 1:** (a) illustrates the 12-core systematic standard protocol, which is defined on a schematic coronal plane of the prostate. (b) Coronal sector definition for the accuracy study (B=Base, M=Mid-Gland, A=Apex, L=Lateral, P=Parasagittal).

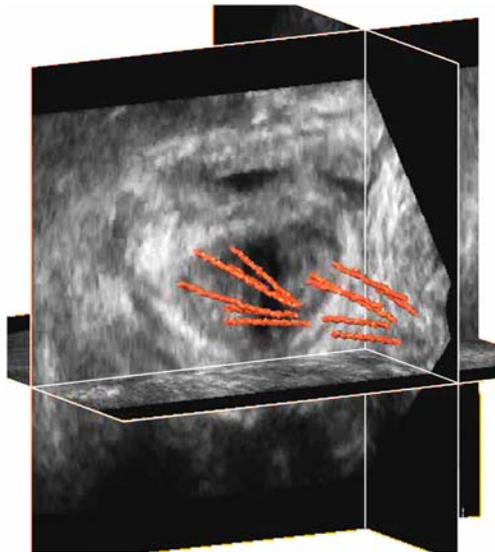

**Figure 2:** 3D biopsy distribution in a reference volume.

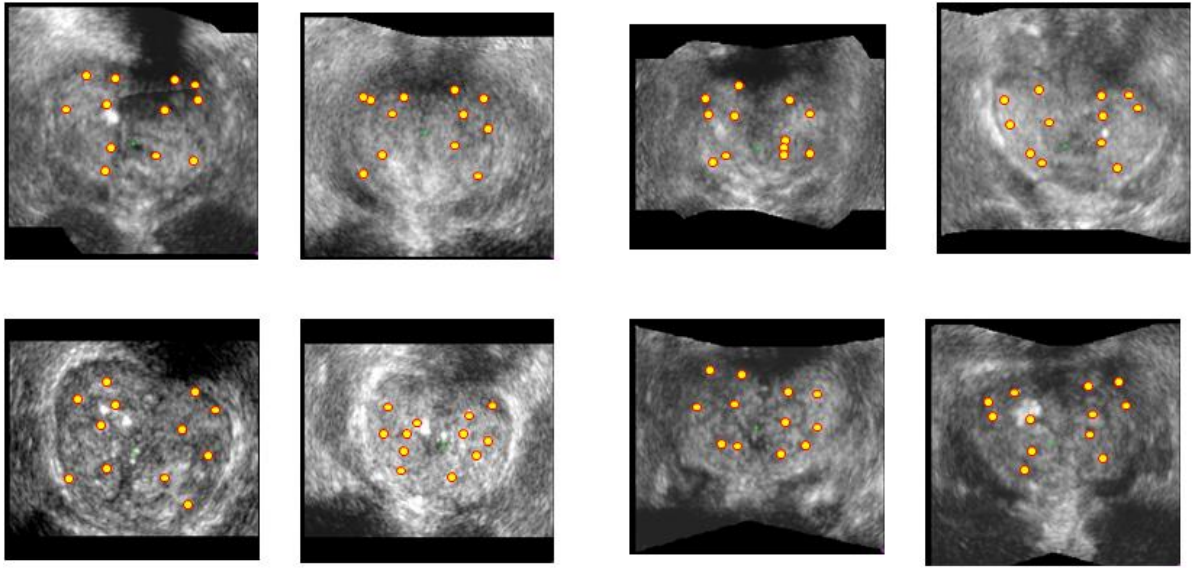

**Figure 3:** Examples of biopsies distribution in the coronal plan for 8 patients.

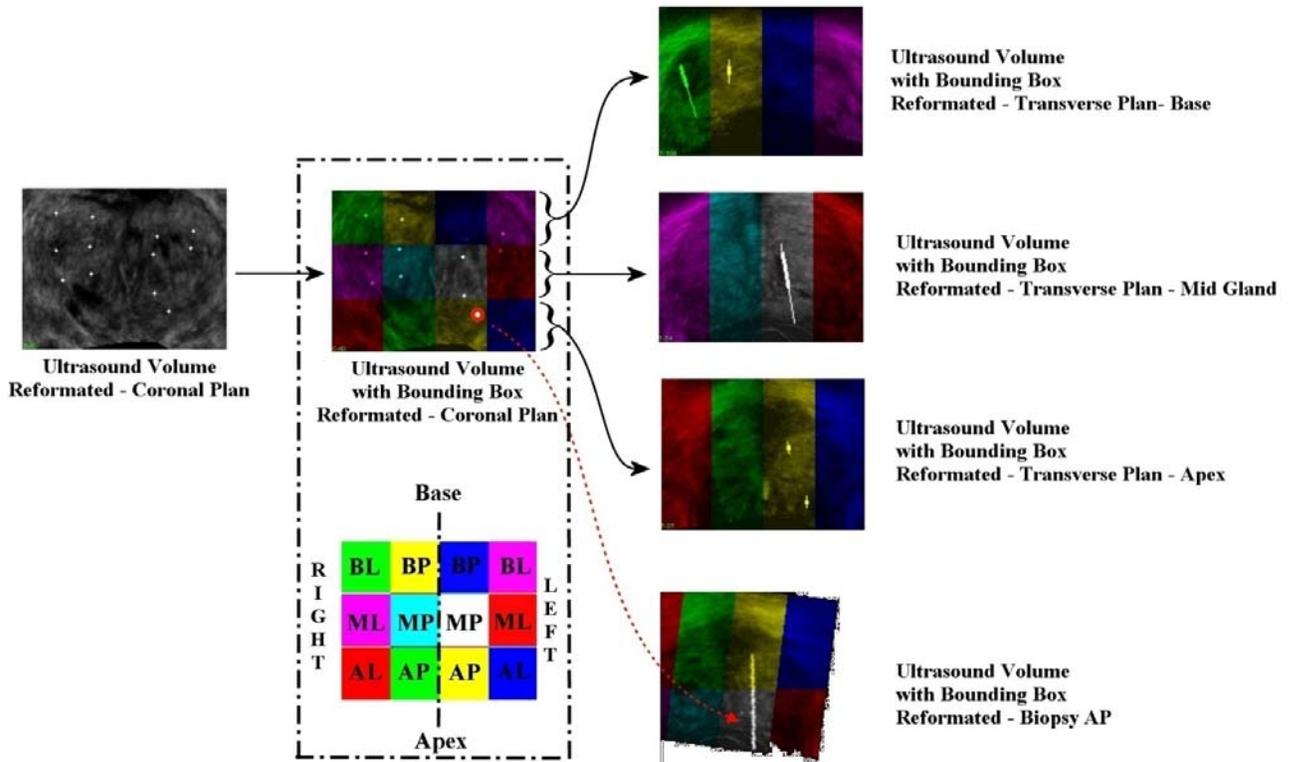

**Figure 4:** Biopsies and reference volume

## *Results:*

- Registration:

The registration method was validated on 237 3D images acquired during biopsy of 14 different patients with an average error < 1.44 mm and a max error 3.84mm. Registration between each volume was computed in 6 seconds. The success rate of all the registration was 96.7% (371 good registrations on 384 volumes).

- Operator accuracy and learning curve:

Table 1 and **Figure 5** show the ratio of biopsies reaching their target and their inner length. On average, the operator reached the target in 67% of all cases. The ratio decreases as the planning approaches the boundaries of the prostate. If we split this data in two equal sets in a chronological order, the success rate is 60% on the first 16 patients and 72% on the last 16 patients. It seems that a learning curve can be highlighted (test for independence of all factors: Chi2 = 5.89, p-value = 0.01523).

| Target | | # of biopsies | % (#) of biopsies inside the target | Biopsy length inside the target (mm) |
|---|---|---|---|---|
| Base Lateral (BL) | Right | 33 | 70% (23) | 14 |
| | Left | 31 | 55% (17) | 12 |
| Base Sagital (BS) | Right | 31 | 65% (20) | 15 |
| | Left | 32 | 66% (21) | 14 |
| Mid Lateral (ML) | Right | 32 | 81% (26) | 15 |
| | Left | 30 | 77% (23) | 15 |
| Mid Sagital (MS) | Right | 32 | 100% (32) | 15 |
| | Left | 31 | 90% (31) | 16 |
| Apex (AL+AS) | Right | 60 | 52% (31) | 12 |
| | Left | 59 | 46% (27) | 13 |
| **Sum/Average** | | **371** | **67% (248)** | **14** |

**Table 1:** Targeting accuracy evaluation result

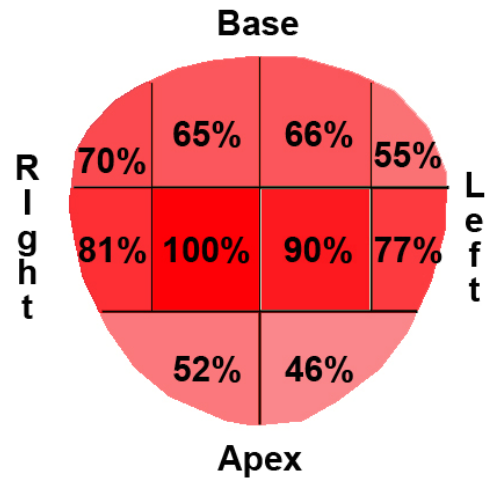

**Figure 5:** Targeting accuracy evaluation result

## *Discussion:*

### *Registration:*
In this study, the registration process does not take into account the deformation induced by the TRUS probe. Nevertheless, this system can be used for quality control, with an accuracy of few millimeters, if the operator is well aware of the deformations.

Moreover, this framework can be used to register data acquired at different time to check if a previous suspicious target has been hit (Figure 6).

### *Operator accuracy and learning curve:*
The low percentage of reached targets and the low inner lengths of sampled tissue inside the targets tend to prove the inadequacy between a theoretical planning pattern and the actual constrained transrectal access. At lateral base and apex sides the lowest ratios may be explained by the difficulty to target and by the low prostate presence in those sectors.

For the first time, at our knowledge, this study allows to quantify a learning curve based on spatial distribution of 2D TRUS prostate biopsies.

*Future:*

Work is in progress to estimate deformations during the procedure. Some preliminary results show that we can estimate theses deformations but the accuracy of such approach is still unknown (see Figure 6).

With 6 seconds per registration the algorithm should be fast enough to achieve frame-rates of 5Hz when making advantage of the massive parallelization capacities provided by modern high-end computers. Then, this method seems robust enough to allow continuous tracking. It could be possible to select a target in the reference volume and guide the clinician to reach it. Work is in progress to register MRI and 3D TRUS reference volumes. This tool should allow to select the target in MRI images and tracking it by this framework. Note that this approach does not rely on any external optical or magnetic tracking system, which means that no cumbersome additional hardware is required.

A good way to enhance the accuracy of this framework could be to hold the TRUS probe with a robot as show on Figure 7.

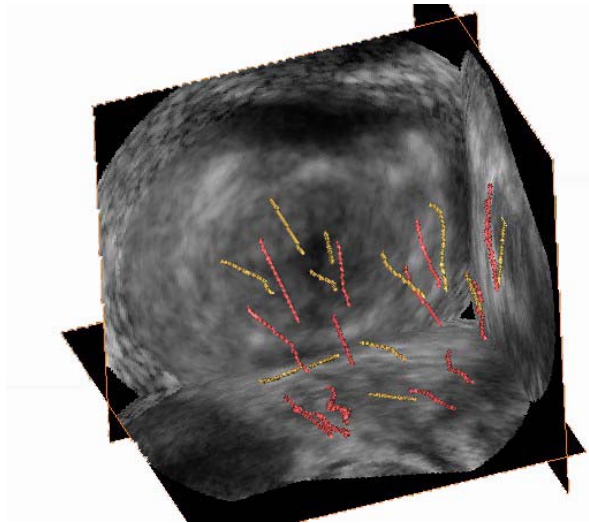

**Figure 6:** 2 sets of biopsies in one reference volume (with deformation)

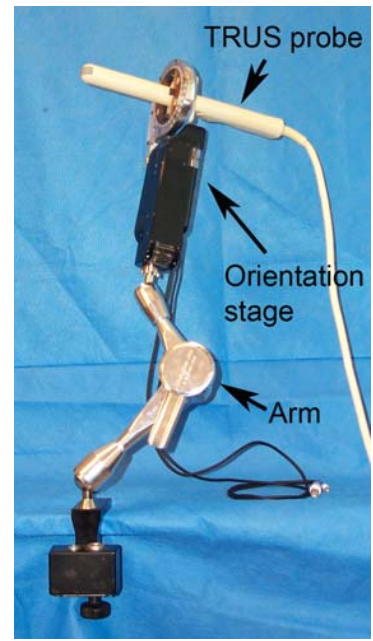

**Figure 7: TRUS robot (From Urobotics Lab - JHU)**

*Conclusion:*

This study shows that it is difficult to accurately reach targets in the prostate using a 2D TRUS probe. On average, the operator reached the target in 67% of all cases and it seems that a learning curve can be highlighted.

Currently, this research framework can be used as a biopsy quality control tool. Moreover, different biopsies sessions can be register to verify that a previous suspicious area has been sampled. Finally, use of a robot in the loop could allow to increase the accuracy and reproducibility.